%% file: neurips_2022.tex
\documentclass{article}

\usepackage[final]{neurips_2022}

\usepackage[utf8]{inputenc} 
\usepackage[T1]{fontenc}    
\usepackage{hyperref}       
\usepackage{url}            
\usepackage{booktabs}       
\usepackage{amsfonts} 
\usepackage{array}
\usepackage{nicefrac} 
\usepackage{chemformula}
\usepackage{microtype}      
\usepackage{xcolor}         
\usepackage{amsmath}
\usepackage{graphicx}
\usepackage{subfigure}

\title{MatKG: The Largest Knowledge Graph in Materials Science - Entities, Relations, and Link Prediction through Graph Representation Learning}

\author{%
  Vineeth Venugopal\\
  Massachusetts Institute of Technology\\
  \texttt{vineethv@mit.edu} \\
\And
  Sumit Pai\\
  Accenture Labs, Dublin\\
  \texttt{sumit.pai@accenture.com} \\
\And
  Elsa Olivetti\\
  Massachusetts Institute of Technology\\
  \texttt{elsao@mit.edu} \\
}

\begin{document}
\maketitle
\input{abstract.tex}
\input{introduction.tex}
\input{methods.tex}
\input{results.tex}

\input{Broaderimpact}
\bibliographystyle{unsrtnat}
\bibliography{references}
\input{Appendix}
\end{document}

%% file: abstract.tex
\begin{abstract}
This paper introduces MatKG, a novel graph database of key concepts in material science spanning the traditional material-structure-property-processing paradigm. MatKG is autonomously generated through transformer-based, large language models and generates pseudo ontological schema through statistical co-occurrence mapping. At present, MatKG contains over 2 million unique relationship triples derived from 80,000 entities. This allows the curated analysis, querying, and visualization of materials knowledge at unique resolution and scale. Further, Knowledge Graph Embedding models are used to learn embedding representations of nodes in the graph which are used for downstream tasks such as link prediction and entity disambiguation. MatKG allows the rapid dissemination and assimilation of data when used as a knowledge base, while enabling the discovery of new relations when trained as an embedding model. 
\end{abstract}

%% file: introduction.tex
\section{Introduction}
Comprehensive knowledge of a given material requires the integration of disparate streams of
information that include compositional data, thermodynamic parameters, applications, phase/symmetry
labels, synthesis and processing routines, as well as physical, chemical, thermal, optical, and functional
properties\cite{william1989structure}. In general, it
is difficult to find all this information in one place, with the result that comprehensive knowledge of a
given material is often missing, even when the data are available. Given the rate at which new data are being
accumulated, the amount of
available data is far greater than what can be accessed or assimilated.
The standard paradigm of data sharing and storage - through peer reviewed scientific
publications and relational databases - remains inadequate for the Materials Genome Age where
artificial intelligence is increasingly employed to accelerate materials discovery and design \cite{tshitoyan2019unsupervised,dima2016informatics,de2019new,de2014materials}. The task of data organization has been approached through custom ontologies that build
relations between data points through manual expert input. While several domain specific ontologies such as Nanomine\cite{mccusker2020nanomine}, Chemos\cite{roch2018chemos}, etc. have been
written over the years, no field-wide ontology exists focused on materials science. Given the onerous task of assigning a relation among individual pairs of data, even highly generalizable
ontologies such as SKOS\cite{miles2009skos} have not been applied to materials at scale.

In this paper, we introduce MatKG, a novel graph database that links major conceptual entities in the discipline using transformer-based large language models. The database is autonomously extracted from over 4 million papers on the topic of materials and includes chemistry, structure, property, application, synthesis, and characterization data that are
aggregated in the form of relational triples <subject, predicate, object>. MatKG has over 2 million unique
relationships among over 80,000 unique entities.

%% file: methods.tex
\section{Methods}

\textbf{Entity Generation}: A Named Entity Recognition (NER) \cite{nadeau2007survey} model was used to extract 80,000 unique entities from the abstracts
and figure captions of over 4 million scientific publications \cite{kim2017virtual}  in the field of material science. Being
information dense, these contain low ‘noise’ and are hence particularly suitable for large scale autonomous data mining\cite{venugopal2021looking,venugopal2019picture,venugopal2021artificial}. The NER model follows the scheme developed in MatScholar \cite{weston2019named} and
is built on MatBERT \cite{weston2019named}, a Large Language Model (LLM) trained on a material science text corpus that
classifies text tokens into one of the following seven categories: Material (CHM), Property (PRO),
Application (APL), Synthesis Method (SYN), Characterization Method (CMT), Descriptor (DSC), and
Symmetry/Phase Label (SPL). Derived from the traditional structure-property-processing-application
paradigm in material science\cite{william1989structure}, these entities encapsulate the sum total of the knowledge of any given
concept, be it a particular chemistry, process, property, or application. Where possible, each entity is
linked to an identifier in Wikipedia  using procedure developed in \cite{spitkovsky2012cross} or the corresponding descriptor page in the Materials
Project\cite{jain2013commentary}. This allows the mapping of entities to broader knowledge bases 
such as DBpedia\cite{auer2007dbpedia} and YAGO\cite{suchanek2007yago}, thereby allowing holistic integration of MatKG with the larger knowledge
graph community.

\textbf{Link Generation} : If entities $e_1$ and $e_2$ have the NER tags $T[e_1]$ and $T[e_2]$, they are assigned the relationship $T[e_1]\_T[e_2]$
and the weight $v(e_1, e_2)$ according to the method detailed in Appendix 5.1. Subsequently, they are either filtered based on a predefined threshold to form knowledge triples of the form <$e_1$, $T[e_1]$\_$T[e_2]$ , $e_2$> (1) or as a quartet of the form <$e_1$, $T[e_1]$\_$T[e_2]$ , $e_2$, $v(e_1,
e_2)$> (2) (See Appendix 5.1). (1) allows the extraction of 160,000 high fidelity links between about
12,000 unique entities, while (2) results in 2 million relations from up to 80,000 unique entities, thereby demonstrating
that a weighted link extraction approach captures far more data - increasing the scope of the
knowledge base.

\textbf{Graph Representation Learning} : The vector representations for the entities in the graph are learnt using knowledge graph embedding models (KGE)\cite{bordes2013translating},\cite{yang2014embedding}, \cite{trouillon2016complex}. The models are evaluated
using mean reciprocal rank (MRR) and hits@(1,10,100) metrics on the test set as described in KGE
literature \cite{cai2018comprehensive}. All models are implemented using the publicly available AmpliGraph Library\cite{ampligraph}.
The model with the highest MRR on the test set was used to perform downstream tasks that are described later. 

%% file: results.tex
\section{Results}
\subsection{Knowledge base creation} 
The autonomously created highly interconnected knowledge graph for materials consists of the seven NER categories and 49 relations (including inverse relations such as $APL\_PRO$ and $PRO\_APL$). The KG is thus a bidirectional digraph. The three most common types of entities are $PRO$, $CHM$, and $CMT$, while the most frequent relations are $CHM\_PRO$, $PRO\_DSC$, and $CHM\_CMT$ (see Appendix, Table 1, 2). The large number of material-property (108 k) and material-application (89 k) triples could correspond to the type of information usually present in abstracts, while characterization related information originate from figure captions. Many papers in the corpus relate to inorganic synthesis\cite{kim2017virtual} which explains the high number of $SMT\_CHM$ (80 k) and $SMT\_PRO$ (67 k) relations. 

\begin{figure}
    \centering
    \includegraphics[scale=0.5]{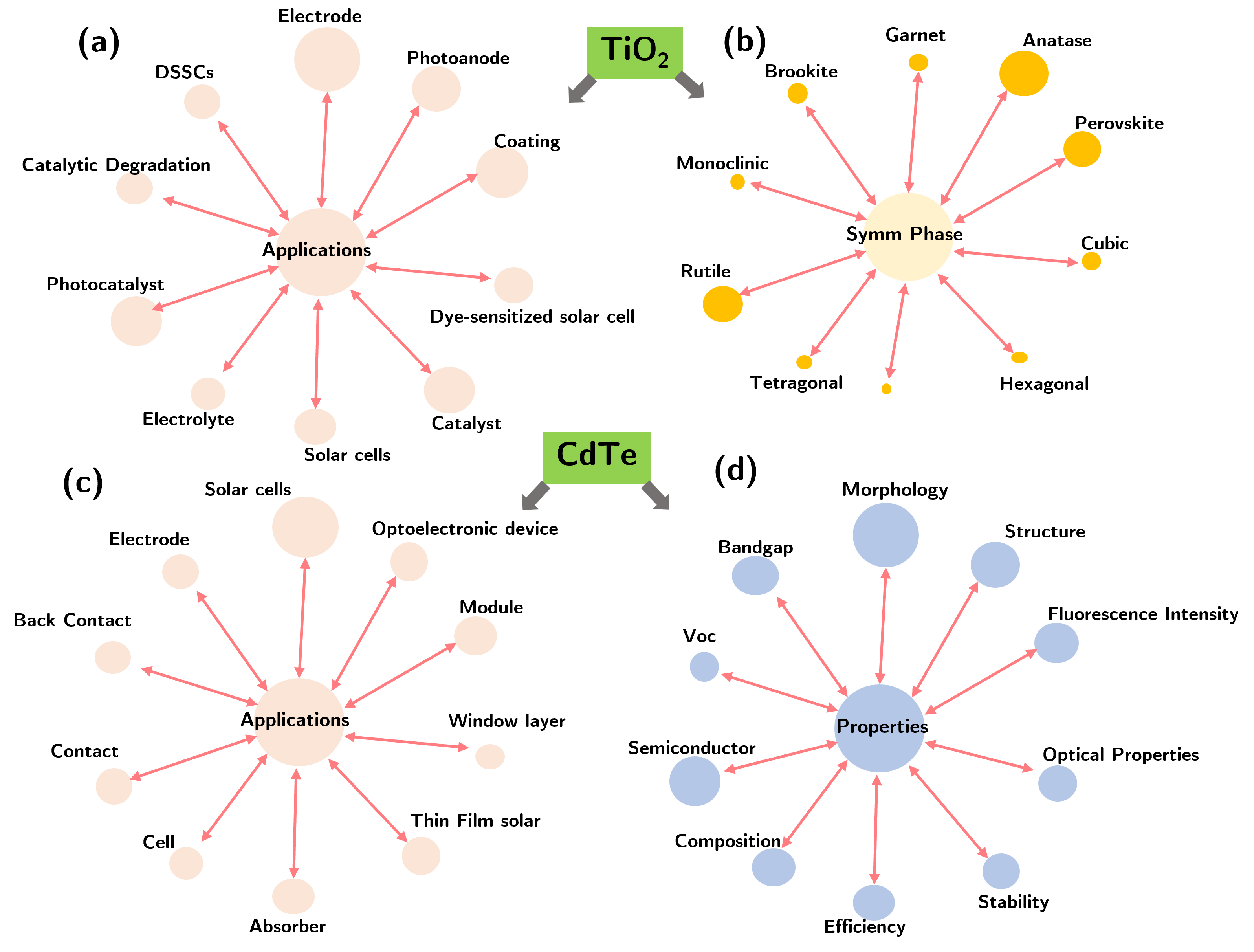}
    \caption{(a) Applications and (b) Symmetry Phase Labels of \ch{TiO2}. (c) Applications and (d) Properties of \ch{CdTe}. The size of the node is proportional to the co-occurence frequency of the link.} \label{graph_vis}
\end{figure}

Together, the acquired data allows the extraction of subgraphs corresponding to wildcard triples such as <\ch{TiO2}, $CHM\_PRO$, ?>, which correspond to the customized query: “what are the properties of \ch{TiO2}?”. Further, by accounting for the co-occurrence frequency, a confidence score can be assigned to each triple as is visually represented in Fig \ref{graph_vis}(a, b)  where the applications and phase labels of \ch{TiO2} are separately extracted and presented as individual bipartite graphs such that the size of the node is  proportional to $v(\ch{TiO2}, e)$. We see that the most common symmetry/phase labels associated with \ch{TiO2} are ‘rutile’ and ‘anastase’, while the most frequent applications are as electrodes, catalyts and for coating. These are in agreement with the widely available literature on the material\cite{guo2019fundamentals}. There is much less information on \ch{CdTe} by comparison (18153 vs 1500 triples), but Fig \ref{graph_vis}(c, d) extracted from MatKG still enables a high-level understanding with some specificity, such as the knowledge that CdTe is used in solar cells and electrodes, and is an optical material as deduced from its properties\cite{shin1983characterization}. 
\begin{figure}[t]
    \centering
    \includegraphics[scale=0.5]{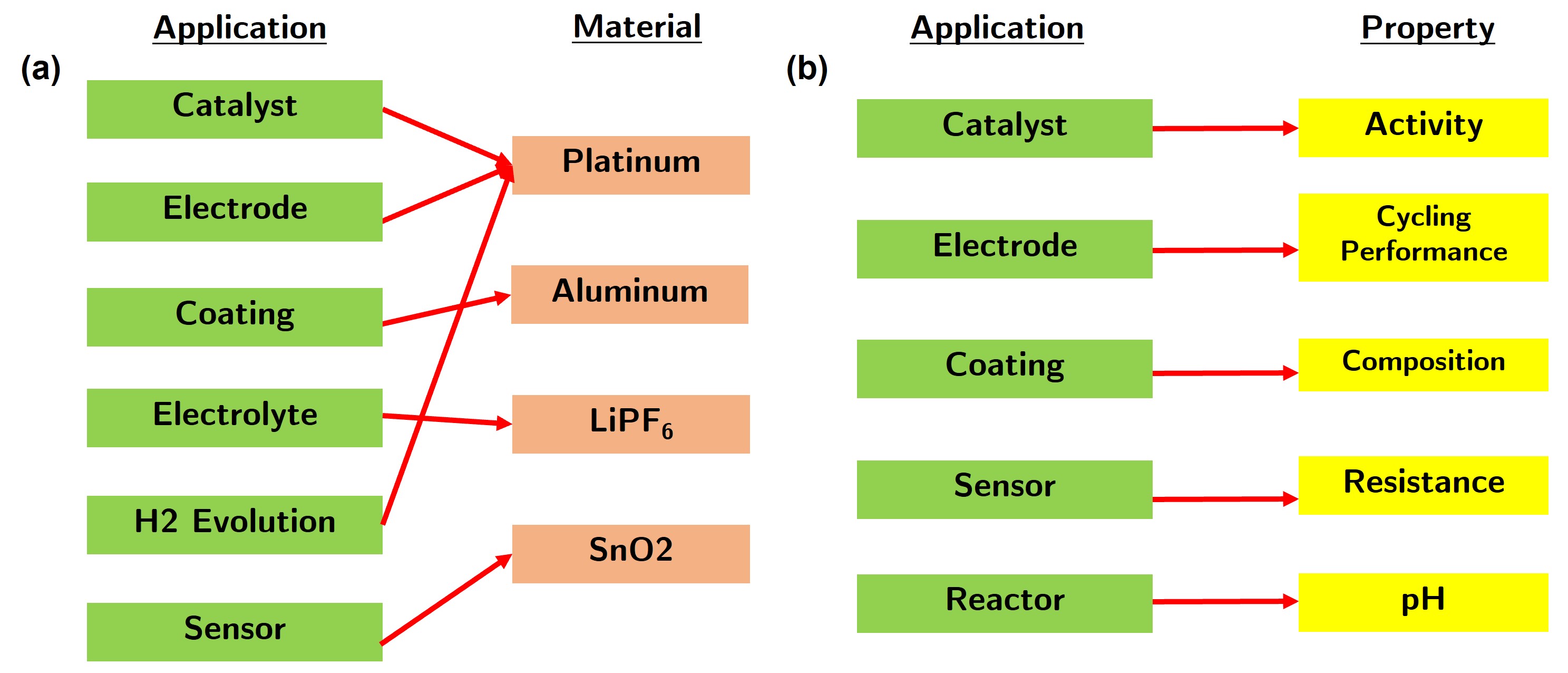}
    \caption{Partitioned (a) application-material (b) application-property subgraphs showing the highest weighted material and property for some select applications}
    \label{Partitioned_Subgraph}
\end{figure}

In addition to material specific queries, MatKG can be partitioned into relation specific subgraphs such as the application-material and application – property graphs in Fig \ref{Partitioned_Subgraph} (a-b), which shows the highest weighted material and property respectively for some select applications. Platinum, perhaps the most widely used metallic catalyst, appears with both ‘catalyst’ and 'hydrogen evolution'. Aluminum is identified as a coating material while \ch{LiPF6} is seen to be an electrolyte, both of which are well known applications of each respectively. In Fig \ref{Partitioned_Subgraph} (b) the most common property associated with electrodes is 'cycling performance', while that of catalyst is ‘activity’. Both are in accordance with our understanding of these concepts. Therefore, MatKG allows the curated visualization and querying of materials specific data directly extracted from literature at unprecedented scale and resolution. 
\subsection{Embedding representation learning}
The TransE\cite{bordes2013translating} model with 150 dimensions is found to have the highest MRR (0.49) on the test set. This model was chosen for discovering new links and for performing entity disambiguation. 


\textbf{Entity Linking}: The similarity between embeddings can be used as a measure of the semantic similarity between entities, in turn becoming a useful tool for both co-reference resolution as well as similar – chemical mapping. As shown in Table 1, several pairs of entities such as ‘qspr’ and ‘quantitative structure property relationship’, or ‘ner’ and ‘net energy ratio’ occupy almost identical positions in MatKG and consequently have very similar graph embeddings. This suggests that they are the same semantic token, even though their lexical distance can be substantial. This form of co-reference resolution is currently not an easy task, especially for the sciences\cite{uzuner2012evaluating}.

\textbf{Link Prediction}: Finally, the KGE model was used to make new link predictions between existing entities in the graph. In this way, the model can be used to discover new applications and properties of existing materials, new properties that can be useful to a given application, or a new characterization method for an existing property, etc. This results in a fuller and more integrated knowledge graph, allowing a holistic analysis of structure-property-processing relations, even when such data is absent in the training literature. 

\begin{table}
\label{similarities-table} \caption{Selected entities and their similarities, demonstrating semantic convergence at the embedding level}
  \centering
  \begin{tabular}{ll}
    \toprule
    \textbf{Entities}     & \textbf{Similarity}      \\
    \midrule
    (qspr, quantitative structure property relationship) & 0.90     \\
    (qmom, quadrature method of moments) & 0.91     \\
    (electromagnetic acoustic resonance, emar) & 0.89     \\
    (ner, net energy ratio) & 0.92     \\
    (let, linear energy transfer) & 0.91     \\
    \bottomrule
  \end{tabular}
    
\end{table}

While the MRR and hits@(1,10,100) are good measures of link predictiveness of the model\cite{khetan2021knowledge}, it is desirable to quantify this inference ability further. To this end, 150 random link predictions were generated by the model across all relationship categories. The top three entities with the highest score for each prediction is manually ranked according to the following criteria: Rank1 if the relationship can be classified as of type\textit{ SKOS: Narrow}, Rank 2 if it is of type\textit{ SKOS: Broad}, and 3 otherwise, where 'narrow' and 'broad' are ontological schema specificed in SKOS \cite{miles2009skos}. An example triple is shown in Table 2, Appendix, which lists the top three model predictions for the applications of \ch{Fe2O3}. Some lithium-ion batteries use lithium-iron-oxide as an electrode, which is usually made by the solid-state reaction of \ch{Li2CO3} and \ch{Fe2O3}, which could explain the first prediction. Since ‘lithium-ion batteries’ is not a direct application of \ch{Fe2O3}, this triple is ranked 2. However, ‘air batteries’ directly use iron/iron oxide as an electrode\cite{requies2013natural} and hence this triple is assigned rank 1. 

Of the 150 x 3 predictions made by the model, 47 \% were found to have a rank 1,  29 \% had a rank of 2, and the rest had a rank of 3 (See Appendix, Table 3 for examples). The utility of this approach is seen in Fig 3, Appendix where previously empty application and characterization subgraphs of Bismuth Telluride (as extracted from MatKG) are populated with meaningful entities through successful link prediction.

%% file: Broaderimpact.tex
\section{Broader Impact}
MatKG is the first step towards the complete synthesis of materials knowledge that allows for the richer databases not just for materials but also for applications, properties, and characterization methods. The ability to predict new links between entities in the graph allows the discovery of new materials for existing applications and properties, in finding new applications of existing materials, and novel correlations between synthesis, characterizations and properties. Consequently, MatKG has broad impact for all the three categories of \textbf{AI-guided materials design}, \textbf{ Automated Synthesis} and for  \textbf{Automated Characterization} . 

%% file: Appendix.tex
\section{Appendix}

\textbf{
\subsection{KG construction}
}

For every entity $e$, the lexical frequency $L(e)$ is defined as the fraction of documents where $e$ is present
at least once, where a document could either be  one of   $N_a$ abstracts or $N_c$ figure captions. For every pair
of entities $(e_1, e_2)$ in a given document, a co-occurrence function $CO(e_1, e_2)$ is defined such that:
\begin{equation}
CO(e_1, e_2) =
\begin{cases}
 1 & \mbox{if both $e_1$ and $e_2$ present in the document} \\
 0 &  \mbox{otherwise}
\end{cases}
\label{eq:alpha}
\end{equation}
The co-occurrence frequency $v(e_1, e_2)$ is then defined as :
\begin{equation}
    v(e_1, e_2) =\frac{ \sum^{N_a + N_c} CO(e_1, e_2)}{N_a + N_c}   
\end{equation}
$v(e_1, e_2)$ therefore is a measure of how many times the given pair of entities $(e_1, e_2)$ co-occur in the
document corpus. Subsequently, two approaches are employed to assign a link to $(e_1, e_2)$.
Approach (1) is based on the premise that if $v(e_1,e_2) > \frac{L(e_1) * L(e_2)}{(N_a + N_c)^{2}}$, then the entities $e_1$ and $e_2$ are
strongly correlated as they occur far more often than their conditional probabilities allow. Approach (2) however, retains all entity pairs but appends their co-occurrence frequency as a weight in the
knowledge representation model

\begin{table}
   
   \centering
   \caption{ NER Categories in MatKG and the number of unique entities in each category.}
   
   \vspace{0.25cm}
   
   \begin{tabular}{lclc}
  
    \toprule
    \textbf{NER Category}     & \textbf{Number of Entities}      \\
    \midrule
     Property (PRO) & 27048     \\
    Chemical (CHM) & 23438     \\
    Characterization Method & 10908     \\
    Synthesis Method & 8547     \\
    Application & 7009     \\
    \bottomrule
    
  \end{tabular}
 \end{table}
  
\vspace{1.5cm}
  
 \begin{table}
   \centering

   \caption{Selected relationships and their instance count in MatKG.}
   
   \vspace{0.25cm}
   
   \begin{tabular}{lclc}
  
    \toprule
    \textbf{Relationship}     & \textbf{Number of Triple}      \\
    \midrule
     $CHM\_CHM$ & 499994     \\
     $PRO\_PRO$ & 368381     \\
     $CHM\_PRO$ & 252714     \\
     $PRO\_DSC$ & 146929     \\
     $CMT\_CHM$ & 141955     \\
     $CHM\_DSC$ & 139740     \\
     $CMT\_PRO$ & 108233     \\
     $CMT\_CMT$ & 100675     \\
     $APL\_PRO$ & 91466     \\
     $CHM\_APL$ & 89117     \\
     $CHM\_SMT$ & 80349     \\
    \bottomrule
    
  \end{tabular}
 
 \end{table} 
 
  \vspace{0.5cm}
  
\begin{table}
  \centering
  \caption{Model predictions for the triple <\ch{Fe2O3}, $CHM\_PRO$, $X$> where $X$ is a property. The triples are ranked according to the scheme described in Results}
  \centering
  \begin{tabular}{llll}
    \toprule
    \textbf{Subject}     & \textbf{relationship}  
    & \textbf{Object}     & \textbf{Rank}  \\
    \midrule
    \ch{Fe2O3}	& $CHM\_APL$	 &$lithium\  ion\ batteries$	&2\\
    \ch{Fe2O3}	& $CHM\_APL$	 &$electrocatalyts$	&1\\
    \ch{Fe2O3}	& $CHM\_APL$	 &$air \ batteries$	&1\\
    
    \bottomrule
  \end{tabular}
\end{table}

\vspace{0.5cm}

\vspace{0.5cm}

\begin{table}
  \centering
  
  \caption{Top three model predicted links for selected examples with model score, custom rank, and cited doi\label{example-pred-links}}
 
  \begin{tabular}{|m{0.17\textwidth}| m{0.124\textwidth}|m{0.13\textwidth}|m{0.05\textwidth}|m{0.05\textwidth}|m{0.474\textwidth}|}
  
    \toprule
    \textbf{Subject}& \textbf{relationship}
    & \textbf{Object}  &\textbf{Score} 
    & \textbf{Rank}  & \textbf{Citation url}\\
    
    \midrule
    optical material & $APL\_CHM$ & \ch{In2O3} & 5.5 & 1 & https://en.wikipedia.org/wiki/Indium(III)\_oxide \\
    optical material & $APL\_CHM$ & \ch{CdO} & 5.27 & 1 & https://en.wikipedia.org/wiki/Cadmium\_oxide \\  
    optical material & $APL\_CHM$ & Zinc Oxide & 5.26 & 1 & https://en.wikipedia.org/wiki/Zinc\_oxide \\ 
    anodic electrode & $APL\_CHM$ & Graphite & 3.00 & 1 & 10.1016/j.ensm.2020.12.027\\
    anodic electrode & $APL\_CHM$ & Carbon-fiber & 3.00 & 1 &
    10.1016/C2015-0-00574-3 \\
    anodic electrode & $APL\_CHM$ & \ch{LiClO4} & 2.90 & 2 & 
    https://en.wikipedia.org/wiki/Lithium\_perchlorate \\
    nuclear reactor & $APL\_CHM$ & Beryllium & 7.02 & 1 &
    https://www.energy.gov/ehss/about-beryllium \\
    nuclear reactor & $APL\_CHM$ & Carbide & 6.41 & 2 & 
    https://en.wikipedia.org/wiki/Uranium\_carbide\\
    nuclear reactor & $APL\_CHM$ & Tungsten & 6.38 & 1 &
    10.1016/j.ijhydene.2016.02.019 \\
    smes & $APL\_PRO$ & dmain & 0.34 & 3 &  N/A \\
    smes & $APL\_PRO$ & transmitted current & 0.28 & 1 & https://en.wikipedia.org/wiki/Superconducting \_magnetic\_energy\_storage \\
    smes & $APL\_PRO$ & u11 & 0.20 & 3 & N/A \\
    reverse water gas shift reaction & $APL\_CHM$ & \ch{C6H5OH} & 5.22 & 3 & N/A \\
     reverse water gas shift reaction & $APL\_CHM$ & Naphtha & 5.22 & 3 & N/A \\
    reverse water gas shift reaction & $APL\_CHM$ & diethylether & 4.79 & 3 & N/A \\
    \bottomrule
  \end{tabular}
\end{table}

 \begin{figure}
    \centering
    \includegraphics[scale=0.5]{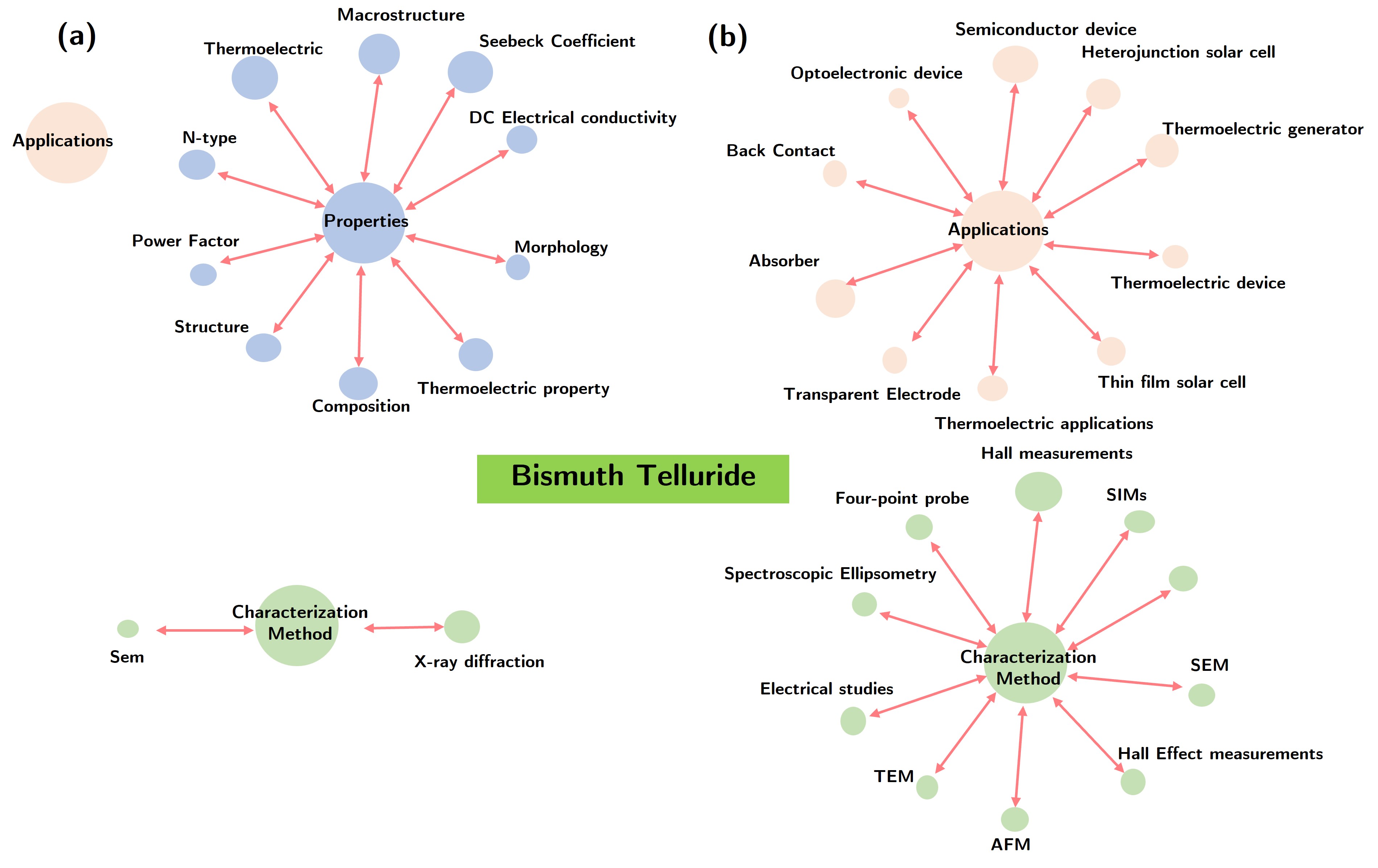}
    \caption{  (a) Original Triples extracted from MatKG and (b) model predicted triples for Bismuth Telluride demonstrating the utility of KGE in complementing material knowledge bases}
    \label{BiTe graph}
\end{figure}

%% file: neurips_2022.bbl
\begin{thebibliography}{28}
\providecommand{\natexlab}[1]{#1}
\providecommand{\url}[1]{\texttt{#1}}
\expandafter\ifx\csname urlstyle\endcsname\relax
  \providecommand{\doi}[1]{doi: #1}\else
  \providecommand{\doi}{doi: \begingroup \urlstyle{rm}\Url}\fi

\bibitem[William and Callister(1989)]{william1989structure}
D~William and J~Callister.
\newblock The structure of crystalline solids.
\newblock In \emph{Materials Science and Engineering}, pages 64--65. Wiley,
  1989.

\bibitem[Tshitoyan et~al.(2019)Tshitoyan, Dagdelen, Weston, Dunn, Rong,
  Kononova, Persson, Ceder, and Jain]{tshitoyan2019unsupervised}
Vahe Tshitoyan, John Dagdelen, Leigh Weston, Alexander Dunn, Ziqin Rong, Olga
  Kononova, Kristin~A Persson, Gerbrand Ceder, and Anubhav Jain.
\newblock Unsupervised word embeddings capture latent knowledge from materials
  science literature.
\newblock \emph{Nature}, 571\penalty0 (7763):\penalty0 95--98, 2019.

\bibitem[Dima et~al.(2016)Dima, Bhaskarla, Becker, Brady, Campbell, Dessauw,
  Hanisch, Kattner, Kroenlein, Newrock, et~al.]{dima2016informatics}
Alden Dima, Sunil Bhaskarla, Chandler Becker, Mary Brady, Carelyn Campbell,
  Philippe Dessauw, Robert Hanisch, Ursula Kattner, Kenneth Kroenlein, Marcus
  Newrock, et~al.
\newblock Informatics infrastructure for the materials genome initiative.
\newblock \emph{Jom}, 68\penalty0 (8):\penalty0 2053--2064, 2016.

\bibitem[de~Pablo et~al.(2019)de~Pablo, Jackson, Webb, Chen, Moore, Morgan,
  Jacobs, Pollock, Schlom, Toberer, et~al.]{de2019new}
Juan~J de~Pablo, Nicholas~E Jackson, Michael~A Webb, Long-Qing Chen, Joel~E
  Moore, Dane Morgan, Ryan Jacobs, Tresa Pollock, Darrell~G Schlom, Eric~S
  Toberer, et~al.
\newblock New frontiers for the materials genome initiative.
\newblock \emph{npj Computational Materials}, 5\penalty0 (1):\penalty0 1--23,
  2019.

\bibitem[De~Pablo et~al.(2014)De~Pablo, Jones, Kovacs, Ozolins, and
  Ramirez]{de2014materials}
Juan~J De~Pablo, Barbara Jones, Cora~Lind Kovacs, Vidvuds Ozolins, and Arthur~P
  Ramirez.
\newblock The materials genome initiative, the interplay of experiment, theory
  and computation.
\newblock \emph{Current Opinion in Solid State and Materials Science},
  18\penalty0 (2):\penalty0 99--117, 2014.

\bibitem[McCusker et~al.(2020)McCusker, Keshan, Rashid, Deagen, Brinson, and
  McGuinness]{mccusker2020nanomine}
James~P McCusker, Neha Keshan, Sabbir Rashid, Michael Deagen, Cate Brinson, and
  Deborah~L McGuinness.
\newblock Nanomine: A knowledge graph for nanocomposite materials science.
\newblock In \emph{International Semantic Web Conference}, pages 144--159.
  Springer, 2020.

\bibitem[Roch et~al.(2018)Roch, H{\"a}se, Kreisbeck, Tamayo-Mendoza, Yunker,
  Hein, and Aspuru-Guzik]{roch2018chemos}
Lo{\"\i}c~M Roch, Florian H{\"a}se, Christoph Kreisbeck, Teresa Tamayo-Mendoza,
  Lars~PE Yunker, Jason~E Hein, and Al{\'a}n Aspuru-Guzik.
\newblock Chemos: orchestrating autonomous experimentation.
\newblock \emph{Science Robotics}, 3\penalty0 (19):\penalty0 eaat5559, 2018.

\bibitem[Miles and Bechhofer(2009)]{miles2009skos}
Alistair Miles and Sean Bechhofer.
\newblock Skos simple knowledge organization system reference.
\newblock \emph{W3C recommendation}, 2009.

\bibitem[Nadeau and Sekine(2007)]{nadeau2007survey}
David Nadeau and Satoshi Sekine.
\newblock A survey of named entity recognition and classification.
\newblock \emph{Lingvisticae Investigationes}, 30\penalty0 (1):\penalty0 3--26,
  2007.

\bibitem[Kim et~al.(2017)Kim, Huang, Jegelka, and Olivetti]{kim2017virtual}
Edward Kim, Kevin Huang, Stefanie Jegelka, and Elsa Olivetti.
\newblock Virtual screening of inorganic materials synthesis parameters with
  deep learning.
\newblock \emph{npj Computational Materials}, 3\penalty0 (1):\penalty0 1--9,
  2017.

\bibitem[Venugopal et~al.(2021{\natexlab{a}})Venugopal, Sahoo, Zaki, Agarwal,
  Gosvami, and Krishnan]{venugopal2021looking}
Vineeth Venugopal, Sourav Sahoo, Mohd Zaki, Manish Agarwal, Nitya~Nand Gosvami,
  and NM~Anoop Krishnan.
\newblock Looking through glass: Knowledge discovery from materials science
  literature using natural language processing.
\newblock \emph{Patterns}, 2\penalty0 (7):\penalty0 100290, 2021{\natexlab{a}}.

\bibitem[Venugopal et~al.(2019)Venugopal, Broderick, and
  Rajan]{venugopal2019picture}
Vineeth Venugopal, Scott~R Broderick, and Krishna Rajan.
\newblock A picture is worth a thousand words: applying natural language
  processing tools for creating a quantum materials database map.
\newblock \emph{MRS Communications}, 9\penalty0 (4):\penalty0 1134--1141, 2019.

\bibitem[Venugopal et~al.(2021{\natexlab{b}})Venugopal, Bishnoi, Singh, Zaki,
  Grover, Bauchy, Agarwal, and Krishnan]{venugopal2021artificial}
Vineeth Venugopal, Suresh Bishnoi, Sourabh Singh, Mohd Zaki, Hargun~Singh
  Grover, Mathieu Bauchy, Manish Agarwal, and NM~Anoop Krishnan.
\newblock Artificial intelligence and machine learning in glass science and
  technology: 21 challenges for the 21st century.
\newblock \emph{International journal of applied glass science}, 12\penalty0
  (3):\penalty0 277--292, 2021{\natexlab{b}}.

\bibitem[Weston et~al.(2019)Weston, Tshitoyan, Dagdelen, Kononova, Trewartha,
  Persson, Ceder, and Jain]{weston2019named}
Leigh Weston, Vahe Tshitoyan, John Dagdelen, Olga Kononova, Amalie Trewartha,
  Kristin~A Persson, Gerbrand Ceder, and Anubhav Jain.
\newblock Named entity recognition and normalization applied to large-scale
  information extraction from the materials science literature.
\newblock \emph{Journal of chemical information and modeling}, 59\penalty0
  (9):\penalty0 3692--3702, 2019.

\bibitem[Spitkovsky and Chang(2012)]{spitkovsky2012cross}
Valentin~I Spitkovsky and Angel~X Chang.
\newblock A cross-lingual dictionary for english wikipedia concepts.
\newblock 2012.

\bibitem[Jain et~al.(2013)Jain, Ong, Hautier, Chen, Richards, Dacek, Cholia,
  Gunter, Skinner, Ceder, et~al.]{jain2013commentary}
Anubhav Jain, Shyue~Ping Ong, Geoffroy Hautier, Wei Chen, William~Davidson
  Richards, Stephen Dacek, Shreyas Cholia, Dan Gunter, David Skinner, Gerbrand
  Ceder, et~al.
\newblock Commentary: The materials project: A materials genome approach to
  accelerating materials innovation.
\newblock \emph{APL materials}, 1\penalty0 (1):\penalty0 011002, 2013.

\bibitem[Auer et~al.(2007)Auer, Bizer, Kobilarov, Lehmann, Cyganiak, and
  Ives]{auer2007dbpedia}
S{\"o}ren Auer, Christian Bizer, Georgi Kobilarov, Jens Lehmann, Richard
  Cyganiak, and Zachary Ives.
\newblock Dbpedia: A nucleus for a web of open data.
\newblock In \emph{The semantic web}, pages 722--735. Springer, 2007.

\bibitem[Suchanek et~al.(2007)Suchanek, Kasneci, and Weikum]{suchanek2007yago}
Fabian~M Suchanek, Gjergji Kasneci, and Gerhard Weikum.
\newblock Yago: a core of semantic knowledge.
\newblock In \emph{Proceedings of the 16th international conference on World
  Wide Web}, pages 697--706, 2007.

\bibitem[Bordes et~al.(2013)Bordes, Usunier, Garcia-Duran, Weston, and
  Yakhnenko]{bordes2013translating}
Antoine Bordes, Nicolas Usunier, Alberto Garcia-Duran, Jason Weston, and Oksana
  Yakhnenko.
\newblock Translating embeddings for modeling multi-relational data.
\newblock In \emph{NIPS}, pages 2787--2795, 2013.

\bibitem[Yang et~al.(2015)Yang, Yih, He, Gao, and Deng]{yang2014embedding}
Bishan Yang, Scott Wen-tau Yih, Xiaodong He, Jianfeng Gao, and Li~Deng.
\newblock Embedding entities and relations for learning and inference in
  knowledge bases.
\newblock In \emph{ICLR}, 2015.

\bibitem[Trouillon et~al.(2016)Trouillon, Welbl, Riedel, Gaussier, and
  Bouchard]{trouillon2016complex}
Th{\'e}o Trouillon, Johannes Welbl, Sebastian Riedel, {\'E}ric Gaussier, and
  Guillaume Bouchard.
\newblock Complex embeddings for simple link prediction.
\newblock In \emph{ICML}, pages 2071--2080, 2016.

\bibitem[Cai et~al.(2018)Cai, Zheng, and Chang]{cai2018comprehensive}
Hongyun Cai, Vincent~W Zheng, and Kevin Chen-Chuan Chang.
\newblock A comprehensive survey of graph embedding: Problems, techniques, and
  applications.
\newblock \emph{IEEE Transactions on Knowledge and Data Engineering},
  30\penalty0 (9):\penalty0 1616--1637, 2018.

\bibitem[Costabello et~al.(2019)Costabello, Pai, Van, McGrath, McCarthy, and
  Tabacof]{ampligraph}
Luca Costabello, Sumit Pai, Chan~Le Van, Rory McGrath, Nicholas McCarthy, and
  Pedro Tabacof.
\newblock {AmpliGraph: a Library for Representation Learning on Knowledge
  Graphs}, 2019.
\newblock URL \url{https://doi.org/10.5281/zenodo.2595043}.

\bibitem[Guo et~al.(2019)Guo, Zhou, Ma, and Yang]{guo2019fundamentals}
Qing Guo, Chuanyao Zhou, Zhibo Ma, and Xueming Yang.
\newblock Fundamentals of tio2 photocatalysis: concepts, mechanisms, and
  challenges.
\newblock \emph{Advanced Materials}, 31\penalty0 (50):\penalty0 1901997, 2019.

\bibitem[Shin et~al.(1983)Shin, Bajaj, Moudy, and
  Cheung]{shin1983characterization}
SH~Shin, J~Bajaj, LA~Moudy, and DT~Cheung.
\newblock Characterization of te precipitates in cdte crystals.
\newblock \emph{Applied Physics Letters}, 43\penalty0 (1):\penalty0 68--70,
  1983.

\bibitem[Uzuner et~al.(2012)Uzuner, Bodnari, Shen, Forbush, Pestian, and
  South]{uzuner2012evaluating}
Ozlem Uzuner, Andreea Bodnari, Shuying Shen, Tyler Forbush, John Pestian, and
  Brett~R South.
\newblock Evaluating the state of the art in coreference resolution for
  electronic medical records.
\newblock \emph{Journal of the American Medical Informatics Association},
  19\penalty0 (5):\penalty0 786--791, 2012.

\bibitem[Khetan et~al.(2021)Khetan, Wetherley, Eneva, Sengupta, Fano,
  et~al.]{khetan2021knowledge}
Vivek Khetan, Erin Wetherley, Elena Eneva, Shubhashis Sengupta, Andrew~E Fano,
  et~al.
\newblock Knowledge graph anchored information-extraction for domain-specific
  insights.
\newblock \emph{arXiv preprint arXiv:2104.08936}, 2021.

\bibitem[Requies et~al.(2013)Requies, G{\"u}emez, Perez~Gil, Barrio, Cambra,
  Izquierdo, and Arias]{requies2013natural}
J~Requies, MB~G{\"u}emez, S~Perez~Gil, VL~Barrio, JF~Cambra, U~Izquierdo, and
  PL~Arias.
\newblock Natural and synthetic iron oxides for hydrogen storage and
  purification.
\newblock \emph{Journal of Materials Science}, 48\penalty0 (14):\penalty0
  4813--4822, 2013.

\end{thebibliography}
